\documentclass{ws-ijbc}
\usepackage{ws-rotating}     
\usepackage{graphicx}
\usepackage{epstopdf}
\begin{document}

\catchline{}{}{}{}{} 

\markboth{Katsanikas and Wiggins}{Phase Space Structure and Transport in a Caldera Potential Energy Surface}

\title{Phase Space Structure and Transport in a Caldera Potential Energy Surface}
\author{MATTHAIOS  KATSANIKAS  AND  STEPHEN  WIGGINS}
\address{School of Mathematics,  University of Bristol, University Walk, Bristol BS8 1TW, UK \\ matthaios.katsanikas@bristol.ac.uk, s.wiggins@bristol.ac.uk
}

\maketitle

\begin{history}
\received{(to be inserted by publisher)}
\end{history}

\begin{abstract}
We study phase space transport  in a  2D caldera potential energy surface (PES) using techniques from nonlinear dynamics.  The caldera PES is characterized by a flat region or shallow minimum at its center surrounded by potential walls and multiple symmetry related index one saddle points that allow entrance and exit from this intermediate region. 
We have discovered four qualitatively distinct cases of the structure of the phase space that govern phase space transport. These cases are categorized according to  the total energy and the stability of  the periodic orbits associated with the family of the central minimum, the bifurcations of the same family, and the energetic accessibility of the index one saddles. In each case we have  computed the invariant manifolds of the unstable periodic orbits of the central region of the potential and the invariant manifolds of the unstable periodic orbits of the families of periodic orbits associated with  the index one  saddles.  The periodic orbits of the central region are, for the first case, the unstable periodic orbits with period 10 that are outside  the stable region of the  stable  periodic orbits of the family of the central minimum. In addition, the periodic orbits of the central region are, for the second and third case, the unstable periodic orbits of the family of the central minimum and for the fourth case  the unstable periodic orbits with period 2 of a period-doubling bifurcation of the family of the central minimum. We have found that there are three distinct mechanisms determined by the invariant manifold structure of the unstable periodic orbits govern the phase space transport. The first mechanism explains the nature of the  entrance of the trajectories from the region of the low energy saddles into the caldera and how they may become trapped in the central region of the potential. The second mechanism describes  the trapping of the trajectories that begin from the central region of the caldera, their  transport  to the regions of the saddles, and the nature of their exit from the caldera. The third mechanism describes the phase space geometry responsible for the dynamical matching of trajectories  originally proposed by Carpenter and described in \cite{col14} for the two dimensional caldera PES that we consider. 
\end{abstract}

\keywords{Chaos and Dynamical Systems, Chemical Reaction Dynamics, Chemical Physics}


\section{Introduction}
\label{intro}
In this paper we analyze the phase space structure and transport in a 2D caldera like potential energy surface (PES).  The caldera PES is characterized by a flat region or shallow minimum at its center  surrounded by potential walls and multiple symmetry related index one saddle points that allow entrance and exit from this intermediate region. This shape of the potential resembles  the collapsed  region within an errupted  volcano (caldera), and this is the reason that Doering \cite{doe02} and co-workers  refer to this type of  potential  as a caldera.

A caldera PES  arises  in many organic chemical reactions, such as the vinylcyclopropane-cyclopentene rearrangement \cite{bal03,gol88}, the stereomutation 
of cyclopropane \cite{dou97}, the degenerate rearrangement  of bicyclo[3.1.0]hex-2-ene \cite{dou99,dou06} or 5-methylenebicyclo[2.1.0]pentane \cite{rey02}.
A key feature of this class of PES is that the stationary point corresponding to the intermediate often possesses higher symmetry than those of either the reactant or product stationary points. Under such circumstances, conventional statistical kinetic models would predict that the symmetry of the intermediate should be expressed in the product ratio. However, in the examples cited this is not observed, and it is the desire to understand this fact that has made analysis of the dynamics of reactions occuring on caldera potentials particularly important \cite{car16,doe68}. Fig. \ref{equi} shows the potential energy contours of a two dimensional caldera studied in Collins et al.  \cite{col14}. In this model the symmetry-related 1-index saddles points provide a configuration space mechanism allowing for the the entrance of trajectories into and exit from the intermediate region of the caldera. Examining Fig. \ref{equi}, one can easily suppose that the caldera is a decision point or crossroads in the reaction where selectivity  between the four points of the saddles, that are indicated by black points, is determined. The subject of this paper is to investigate the phase space structure and transport associated mechanisms of the decision in the caldera and the origin of the selectivity between the four regions of the saddles. In order to address these questions Collins et al.\cite{col14} performed a detailed study of the trajectories in this two dimensional model. Broadly speaking, they identified two types of trajectory behaviour:   \textbf{dynamical matching}   \cite{col14} and trajectory trapping in the caldera.

In this paper, we analyze  the phase space mechanisms and the origin of the trajectory behaviour that was found in \cite{col14}. For this reason we  study the phase space structure of the caldera 
and we use many methods from the theory of dynamical systems in order to have a deep insight of the dynamics of the system. We have computed all of the basic families of periodic orbits of the system and the invariant objects around them. We have found the mechanisms that are responsible for the entrance and the exit from the caldera and the phase space mechanisms for trapping and "dynamical matching". This could not have been achieved without the study of the phase space structure since these mechanisms are based on the geometrical objects in the phase space (invariant manifolds of the unstable periodic orbits).

The structure of the paper is the following. In section \ref{sec.1} we describe  the mathematical model for our system and the method of the surface of section that we used for the study of the phase space structure, and the families of the periodic orbits. In  section \ref{res} we
describe our  results for the four different cases of the structure of the phase space that we discovered in our study. Then we present  the summary and our conclusions in section \ref{sec:summary}. For completeness we give an appendix A describing  how we characterize the linear stability of periodic orbits and their bifurcations and an appendix \ref{divsur} that describes the computation and characterization of periodic orbit dividing surfaces. 

\section{Model and Methods}
\label{sec.1}

In this section we describe the 2D autonomous Hamiltonian system that is a model for the dynamics on the caldera. We introduce the method of Poincar{\'e} surface of section used to investigate the phase space structure of  our Hamiltonian System  and we also present the families of periodic orbits of our system. 

\subsection{Hamiltonian system and equations of motion}
\label{sub.1}
The Caldera potential  has been used in a previous study from \cite{col14}. As we described in the 
\ref{intro}, the topography of the potential is similar with the collapsed region of an erupted volcano. It has at the center a shallow minimum  where there is a stable equilibrium point,
 the central minimum (see Table \ref{ta1}) . 
The shallow minimum of the potential is surrounded by potential walls.  On these potential walls we have the 
existence of four 1-index saddles (two for lower values of Energy and other two for higher values of Energy, see Table \ref{ta1}) that allow the entrance into and exit from the  intermediate region of the caldera. The  four regions of the saddles are the only regions where  molecules can enter  into and exit from the caldera. 
The caldera  has a stable equilibrium point at the center (that is represented by a black point at the center of the  Fig. \ref{equi}) and it is like a decision point in the reaction between four different paths that lead to four different regions of saddles and then to the exit from the caldera (that are represented by four black points in the  Fig. \ref{equi}).  In addition, the entrance into the caldera is achieved through the four different  regions  of the saddles.The potential is:

 \begin{figure}
 \centering
\includegraphics[angle=0,width=8.0cm]{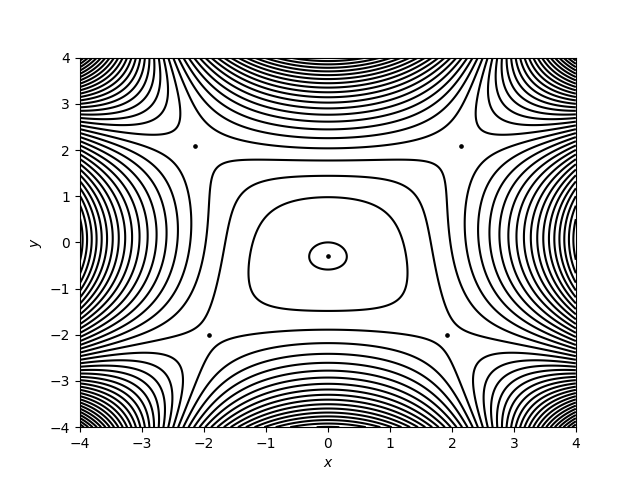}
\caption{The stationary points (depicted by black points)  and contours of the potential.}
\label{equi}
\end{figure}

\begin{eqnarray}
\label{eq1}
V(x,y)=c_1 r^2+c_2 y - c_3 r^4 cos[4 \theta]=
\nonumber\\
c_1(y^2+x^2) + c_2y - c_3(x^4 + y^4 - 6 x^2 y^2)
\nonumber\\
\end{eqnarray}
where $r^2=x^2+y^2$, $cos[\theta]=x/r$. Potential parameters are taken to be $c_1=5$, $c_2=3$,$c_3=-3/10$.
The potential is symmetric with respect to the y-axis. 

The Hamiltonian of the system is:

\begin{eqnarray}
\label{eq2}
 H(x, y, p_x, p_y )=\frac {p_x^2}{2m}+\frac{p_y^2}{2m} + V(x, y)
\end{eqnarray}

with  potential $V(x,y)$  in  eq. \eqref{eq1} and $m=1$. 

The equations of motion are:

\begin{eqnarray}
\label{eq3}
 \dot x= \frac{p_x}{m} \\
 \nonumber\\
 \dot y =\frac{p_y}{m}\\
 \nonumber\\
 \dot p_x=-\frac {\partial V} {\partial x} (x,y)\\
 \nonumber\\
 \dot p_y =-\frac {\partial V} {\partial y} (x,y)
\end{eqnarray}

\begin{table}
\begin{center}
\caption{Stationary points of the potential  given in eq. \eqref{eq1} ("RH" and "LH" are the abbreviations for right hand and left hand respectively)}
\begin{tabular}{l  l  l  l}
\hline
Critical point & x & y & E \\
\hline
Central minimum & 0.000 & -0.297 & -0.448 \\
Upper LH saddle  & -2.149 & 2.0778 & 27.0123 \\
Upper RH saddle  & 2.149  &  2.0778 & 27.0123 \\
Lower LH saddle & -1.923 & -2.003  & 14.767  \\
Lower RH saddle & 1.923  & -2.003 & 14.767 \\
\hline
\end{tabular}
\end{center}
\label{ta1}
\end{table}

\subsection{Surfaces of section and Periodic orbits}
\label{sub2}
In this paper we used the computational method of surfaces of section in order to study the phase space of  a Hamiltonian system. The phase space of an autonomous Hamiltonian system with two degrees of freedom  is  a 4-dimensional space $(x,y,p_x,p_y)$. The energy of the system (the value of the Hamiltonian) is a constant of motion and  one variable, for example $p_y$ (as we did in this paper), can be derived from the other three variables for fixed values of energy. 
This means that the phase space of the system is reduced to a 3-dimensional space $(x,y,p_x)$. If we define a particular surface $S(x,y,p_x,p_y)=const.$ within every fixed interval of time, then the first intersection $S_1$ of an initial point $S_0$   is called the first return of $S_0$ and the surface $S$ as Poincar{\'e} surface of section . The returns of the points on $S$, for a fixed value of energy, define a  map that is called as  Poincar{\'e} map  \cite{poi92,bir60}. The surfaces of section help us to relate  the study of the  phase space of a 2D autonomous Hamiltonian system  to the study of the phase space of a 2D map, the Poincar{\'e} map. This allows us  a simple visualization of the phase space geometry of the system.   In  all cases we used the Poincar{\'e} surfaces of section $y=const$ 
(2D space ($x,p_x$)) with $p_y>0$ that is   computed from the other coordinates for a fixed value of energy. On a 2D Poincar{\'e} map the phase space areas are conserved  \cite{poi92}.

 Every solution of a system like this of \eqref{eq2}  is periodic if  it repeats itself after time T, that is called the period. For example, If we define  a trajectory in the phase space as  $(x(t),y(t),p_{x}(t),p_{y}(t))$ (t is the time), we can  denote the initial conditions of a periodic orbit  as $(x_p(0),y_p(0),p_{x_p}(0),p_{y_p}(0))$ at time t=0. After  time $T$ (the period of the periodic orbit) we will  have that $(x_p(0),y_p(0),p_{x_p}(0),p_{y_p}(0))$ $=(x_p(T),y_p(T),p_{x_p}(T),p_{y_p}(T))$. A periodic orbit is a one-dimensional  closed curve in the phase space. A periodic orbit of a Hamiltonian system of two degrees of freedom is represented by  at least one fixed point on the 2D Poincar{\'e} surface of section before closing. The  periodic orbits  that are represented by more than one point, $n>1$ points, are periodic orbits with high order multiplicity and we will refer to them as periodic orbits with period n or as n-periodic orbits. For example a periodic orbit with period 10 (10-periodic orbit) is represented by 10 points on the  2D Poincar{\'e} section. 
 
According to the theory of linear stability of periodic orbits in autonomous  Hamiltonian systems with two degrees of freedom, we have two types of periodic orbits; the {\bf stable} and {\bf unstable} periodic orbits. In the case of the stable periodic orbits we have the existence of  2-dimensional  invariant tori (KAM invariant tori) surrounding them, according the KAM  theorem \cite{kolmo54,arn63,mos62}. This means that there are  trajectories (quasi-periodic orbits) around the stable periodic orbits that lie forever on these invariant tori.  These KAM invariant tori are topological barriers on the energy surface of the  2D autonomous Hamiltonian systems. As a consequence, trajectories inside the tori are trapped inside forever and trajectories outside the tori can never enter the tori. On the contrary, the unstable periodic orbits repel the trajectories in their neighbourhood.

\begin{figure}
 \centering
\includegraphics[angle=0,width=3.8cm]{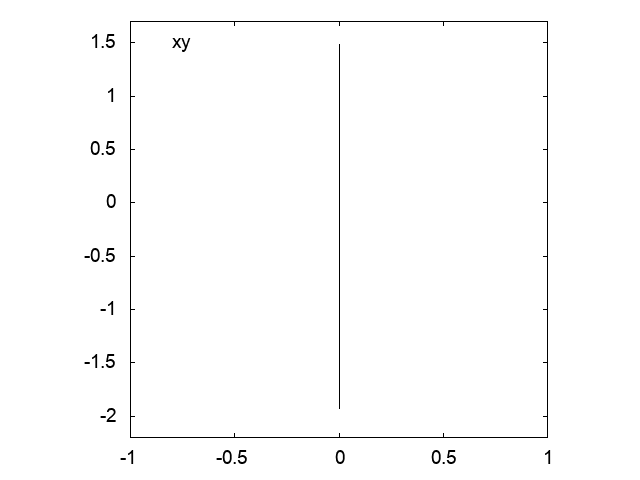}
 \includegraphics[angle=0,width=3.8cm]{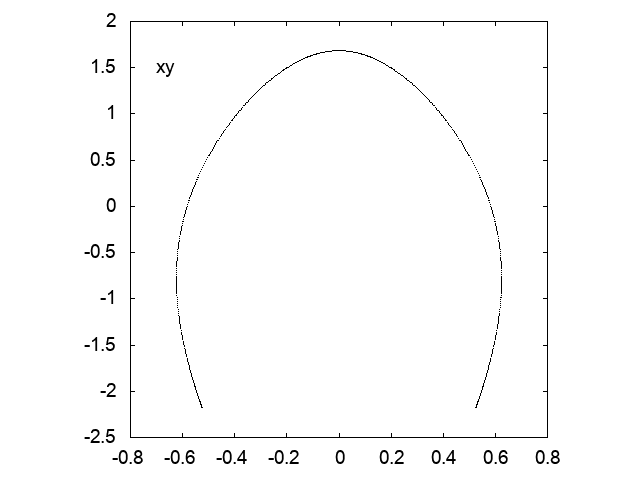}
 \includegraphics[angle=0,width=3.8cm]{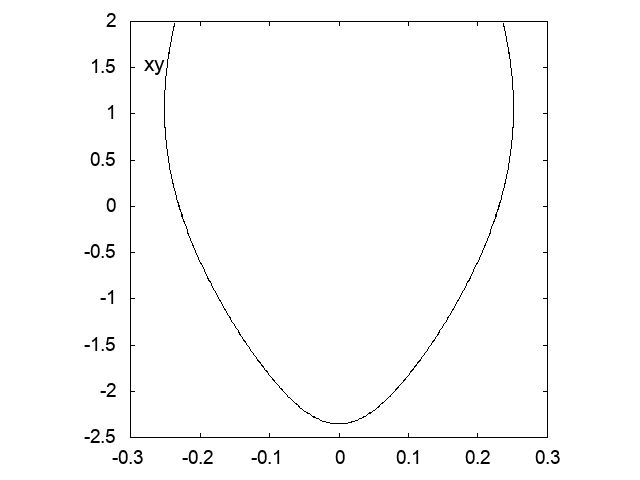}
 \caption{The periodic orbits of the family of the central minimum and its bifurcations in the configuration space. Left panel: The periodic orbit   of the family of the central minimum for E=17. Central panel: The periodic orbit of the first period-doubling bifurcation of the family of the central minimum for E=23. Right panel: The periodic orbit of the second  period-doubling bifurcation of the family of the central minimum for E=30. }
\label{per1}
\end{figure}

\begin{figure}
 \centering
\includegraphics[angle=0,width=5.0cm]{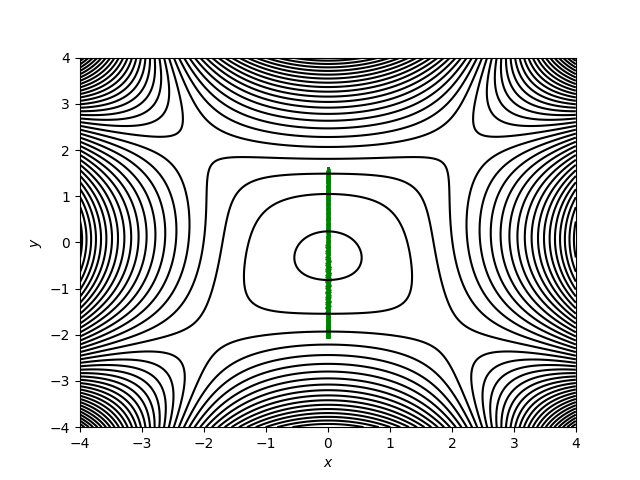}
 \includegraphics[angle=0,width=5.0cm]{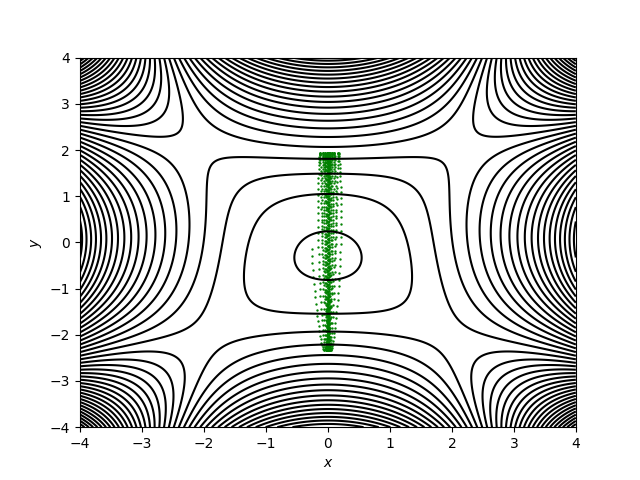}
 \includegraphics[angle=0,width=5.0cm]{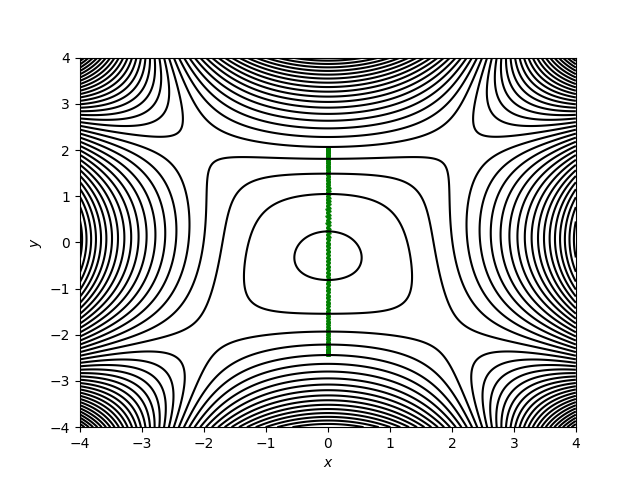}\\
  \includegraphics[angle=0,width=5.0cm]{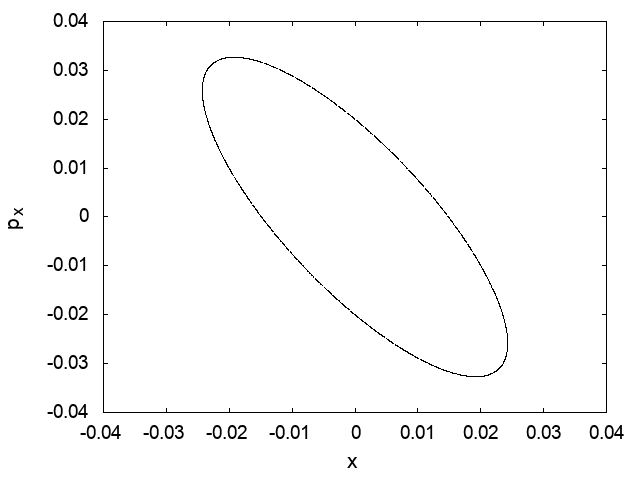}
  \includegraphics[angle=0,width=5.0cm]{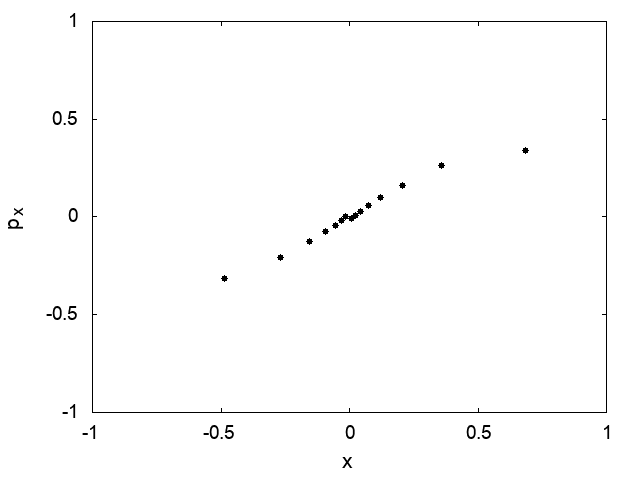}
 \includegraphics[angle=0,width=5.0cm]{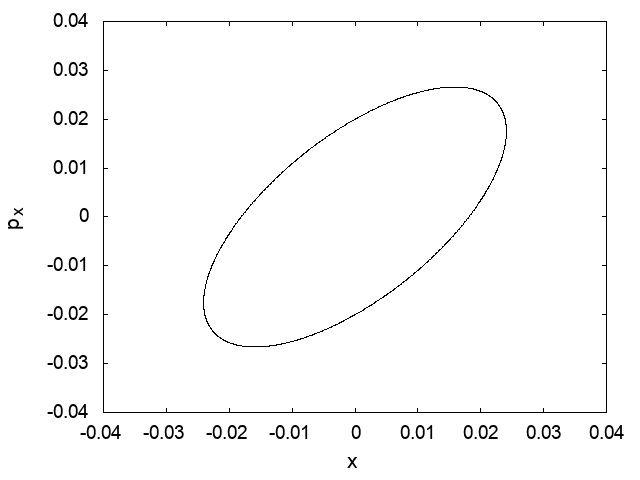}\\
 \caption{The  resulting trajectories if we perturb  (with $\Delta p_x=0.02$  and $t=20$ time units ) the  periodic orbits  of the family of periodic orbits of the central minimum  for values of Energy E=20 (upper left panel), E=29 (upper central panel) and E=33 (upper right panel). We also depict the corresponding Poincar{\'e}  surfaces of section for  E=20 (lower left panel), 
 E=29 (lower central panel) and E=33 (lower right panel).  }
\label{per3}
\end{figure}

We can understand better the different trajectory behaviour in the neighbourhood of the two types of periodic orbits with an example. In this example, we study the trajectory behaviour in the neighbourhood of stable and unstable periodic orbits. This can be achieved if we apply a small perturbation to  the initial conditions of a  periodic orbit. In our example we used the periodic orbits of  the family of the central minimum. The periodic orbits of this family are represented by vertical straight lines in the configuration space (see as example the periodic orbit for E=17 at Fig. \ref{per1}). If we perturb the stable periodic orbits of the family for E=20 and 33 the resulting trajectories are very close to the vertical straight line of the periodic orbit (Fig. \ref{per3}). But if we perturb the unstable periodic orbit for E=29 we observe that the resulting trajectory diverges  from the vertical straight line of the periodic orbit (Fig. \ref{per3}). As we see, the range of the   deviation from the periodic orbit does not depend on the value of the energy but on the kind of linear stability of periodic orbits (stable or unstable). In addition, we observe that the  trajectories for E=20 and 33 are  represented by invariant curves on the surfaces of section (Fig.\ref{per3}). This happens because as we explained above the trajectories  in the neighbourhood of the stable periodic orbits lie on 2-dimensional tori in the phase space that are represented by invariant curves on the 2D surfaces of section. In the case of a trajectory in the neighbourhood of the unstable periodic orbit (for E=29), the trajectory goes far  away from the periodic orbit  and this behaviour is represented on the 2D surfaces of section (Fig. \ref{per3}). The method of the computation and the mathematical description  of the linear stability of periodic orbits is given in appendix A.

\begin{figure}
                      \centering
                  \includegraphics[angle=0,width=8.0cm]{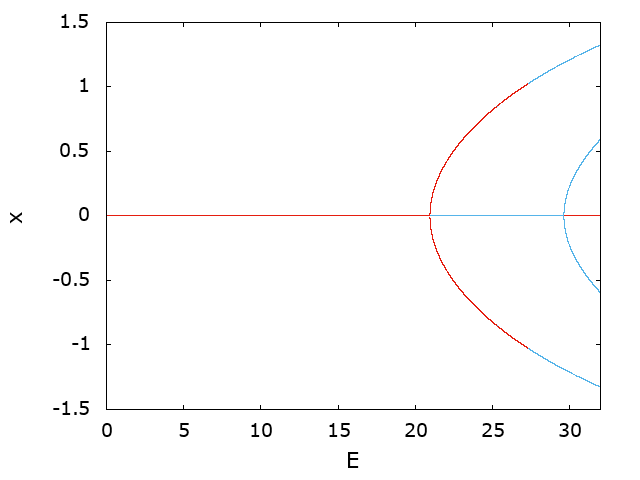}
                  \caption{The diagram of the x-coordinate of the initial conditions on the plane $y=0$, for the   families of periodic orbits that start from the central minimum and their bifurcations, versus energy. The red and cyan parts of the families indicate the stable and unstable parts of the families of periodic orbits, respectively.}
                  \label{char}
                 \end{figure}

 It is  known \cite{poi92} that if we start with a particular periodic orbit of a system and then we vary continuously a parameter of the system we find a set of periodic orbit that is called a family.
 If we apply the Lyapunov subcenter theorem \cite{rab82,wei73,mos76}  at the case of the stable equilibrium point, we have the existence of 
 at least two families of periodic orbits that are associated with the equilibrium point (central minimum). 
We computed the two families that are associated with the central minimum  and their bifurcations for our system. By means of a standard continuation method (using the Newton-Raphson algorithm) we computed the two families on the plane  $y=0$. The   initial conditions $(x,y,p_x,p_y)$, on the plane $y=0$, for the  one family is $(0,0,0,p_y>0)$ and for the other is $(0,0,0,p_y<0)$.  In  Fig. \ref{char}, we give the diagram of  the x-coordinate (on the plane  $y=0$) of these families and their bifurcations  versus energy. We depict the  stable and unstable parts of  the families of periodic orbits with red and cyan color, respectively.  

The families of periodic orbits of the central minimum are initially stable but  then they become unstable, at the value for the energy  E=21 (Fig. \ref{char}). At this point, we have a new supercritical period-doubling bifurcation (see appendix A)  for each of the two families.  This means that we have a   new family of periodic orbits with double period than that  of the parent family. The new  family has two symmetric branches that are represented by two symmetric arches in Fig. \ref{char}.  This family is initially   stable  and then it becomes unstable for larger value of the energy (Fig. \ref{char}).

After the  bifurcation point, the families of periodic orbits of the central minimum become again stable for E=29.64 (see Fig.\ref{char}). At this  point, we have  a new subcritical period-doubling (see appendix A) bifurcation, for each of two families, that is unstable (Fig. \ref{char}).  We noticed that eventually  each of the two families of periodic orbits of the  central minimum  has two period-doubling bifurcation at the points where we have transition from stability to instability and vice versa. The periodic orbits of the family of the central minimum and  their  bifurcations are depicted in the configuration space in  Fig. \ref{per1}.

\begin{figure}
 \centering
\includegraphics[angle=0,width=5.0cm]{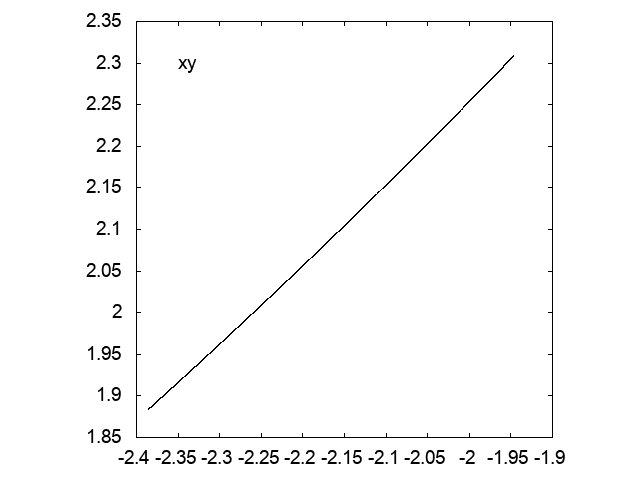}
 \includegraphics[angle=0,width=5.0cm]{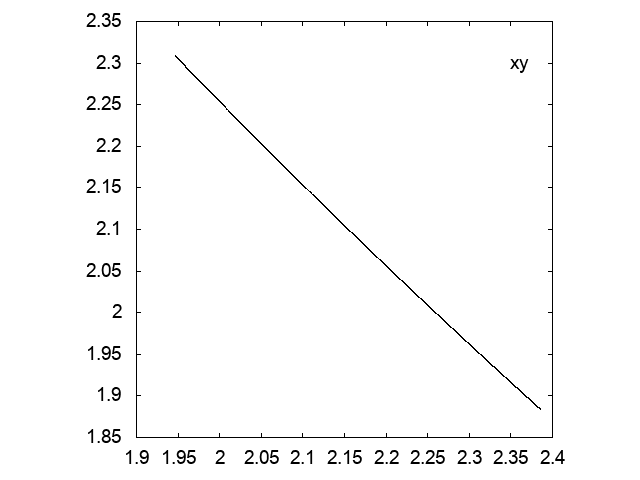}\\
 \includegraphics[angle=0,width=5.0cm]{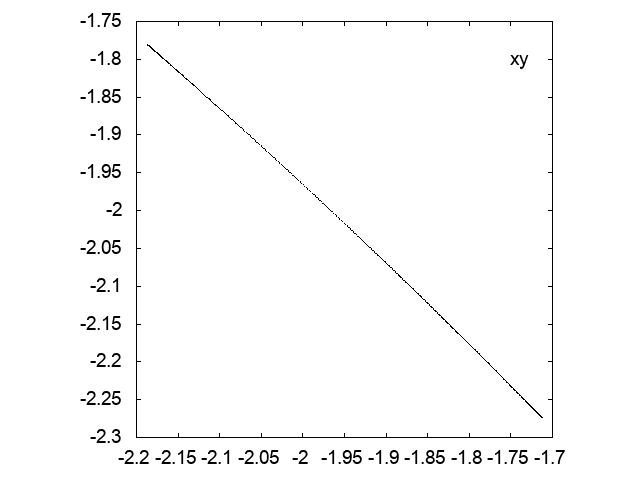}
  \includegraphics[angle=0,width=5.0cm]{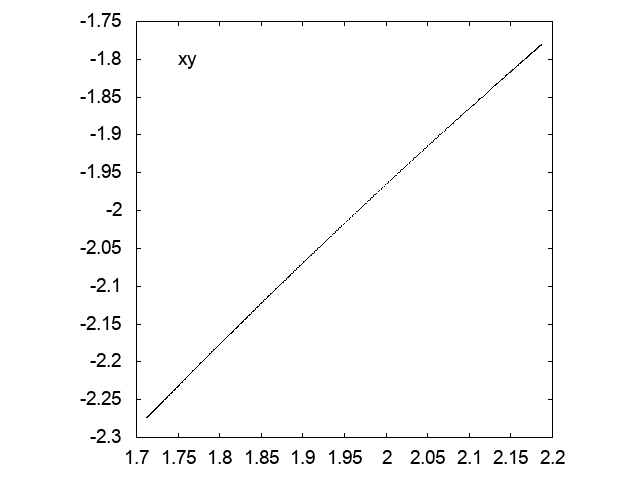}\\
 \caption{The periodic orbits of the families of saddles in the configuration space. Upper left panel: The periodic orbit of the family of the upper left hand saddle for E=29. Upper right panel: The periodic orbit of the family of the upper right hand saddle for E=29. Lower left panel: The periodic orbit of the family of the lower left hand saddle for E=17. Lower right panel: The periodic orbit of the family of the lower right hand saddle for E=17.}
\label{per2}
\end{figure}

According to the Lyapunov subcenter theorem \cite{rab82,wei73,mos76} 
we have at least one family that is associated with every index 1 saddles of our system. In our case we found two families for every  index 1 saddle and we computed them.  We computed these families using a standard continuation method (using the Newton-Raphson algorithm) on the planes $y=a$, with $a=-1.8019693$ for the case of the lower energy saddles and $a=1.884090$ for the case of higher energy saddles. The general form of the initial  conditions $(x,y,p_x,p_y)$ for the  one family  of saddles is $(x_0,a,p_{x_0},p_y>0)$ and for the other is  $(x_0,a,-p_{x_0},p_y<0)$ . In particular, the families  of upper right hand (RH)  saddles  have $x_0>0$ and $p_{x_0}<0$, the families of  upper left (LH) hand saddles have $x_0<0$ and $p_{x_0}>0$, the families  of lower  right hand (RH)  saddles  have $x_0>0$ and $p_{x_0}>0$ and the families of  Lower  left (LH) hand  saddles have $x_0<0$ and $p_{x_0}<0$. These families are unstable. The periodic orbits of the families of the saddles are depicted in the configuration space in  Fig. \ref{per2}.

In all cases, we  studied  the cases of periodic orbits with positive values $p_y$. The results are similar for the case of the periodic orbits with negative values $p_y$. 

\section{Results}
\label{res}
In this section we investigate the  different kinds of trajectories that enter into and exit from the caldera using the notion of dividing surfaces. 
Then we study the phase space structure in order to  understand  the origin of the trajectory behaviour and the mechanisms that support this behaviour. 
For this reason, we use the notion of Poincar{\'e} section and we compute the invariant manifolds of the unstable periodic orbits. 

We encountered four different  cases for the trajectory behaviour and the phase space structure of our model.  These  cases are:

\begin{enumerate}
\item {\bf Case I}:  In this case the periodic orbits of the  families of the central minimum are stable. As we can see in Fig. \ref{char}, 
these families are initially  stable and  continue as stable 
for  values of Energy until  E=21. For these values of energy we have  also the unstable periodic orbits of the families of low energy saddles.  

\item {\bf Case II:} In this case the periodic orbits of the families 
of the central minimum are unstable  after the first  period-doubling 
bifurcation of these families (for values of Energy larger than E=21 - see Fig. \ref{char}).  Moreover,  we consider the interval of energy before  the appearance of high energy saddles (for E=27.0123).  For these values of energy we have  also  the unstable periodic orbits of the families of  low energy saddles.
     
\item {\bf Case III :} In this case the periodic orbits of the families of the  central minimum  are unstable but they exist after the appearance 
 of high energy saddles (for E=27.0123) and before the second period-doubling bifurcation of these families (for E=29.64). For these values of energy we have also the unstable periodic  orbits of the families of  high and low  energy saddles.
   
 \item {\bf Case IV:}  In this case, the periodic orbits of the families of the central minimum  are stable. This case is  after the second period-doubling bifurcation of these families (for E=29.64 - see Fig. \ref{char}).  For these values of Energy we have  again the  unstable periodic  orbits of the families of  high and low energy saddles.
\end{enumerate}

In the following sections, we study  a representative example  for every case. 

\subsection{Case I}
\label{res1a}
\begin{figure}
 \centering
\includegraphics[angle=0,width=5.0cm]{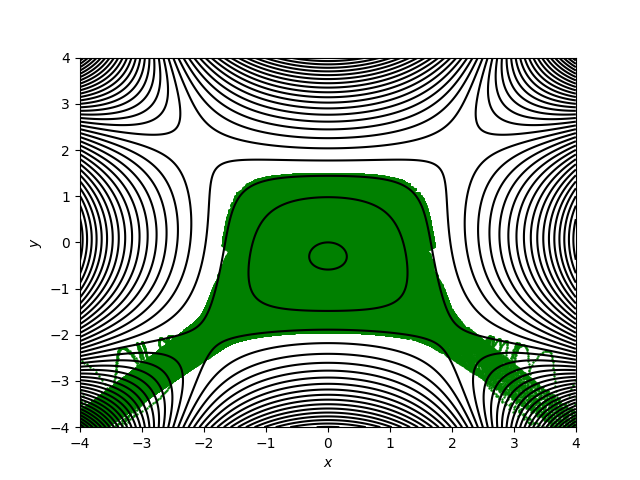}
 \includegraphics[angle=0,width=5.0cm]{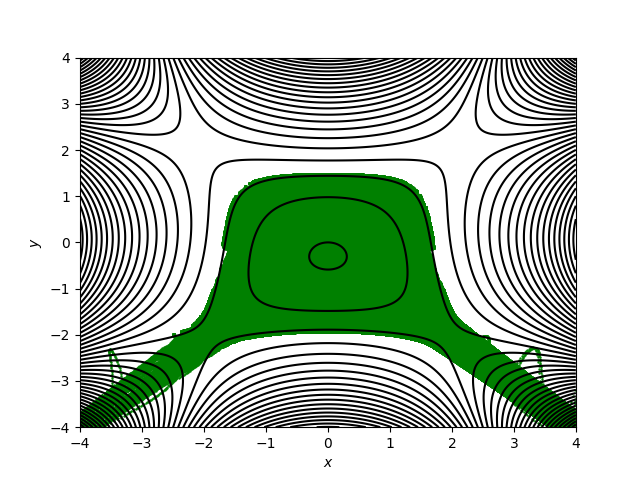}\\
 \includegraphics[angle=0,width=5.0cm]{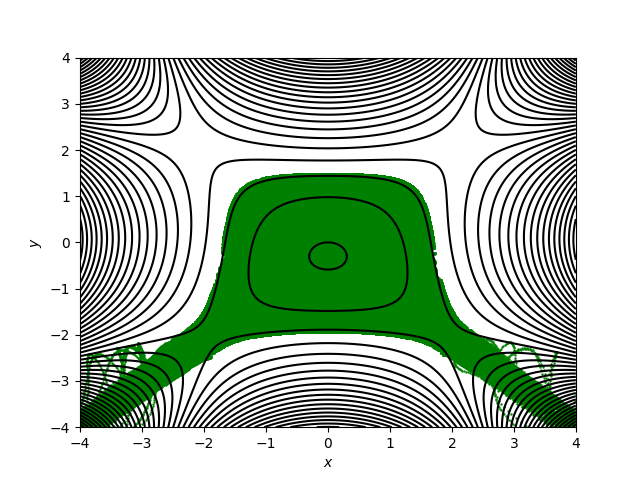}
 \includegraphics[angle=0,width=5.0cm]{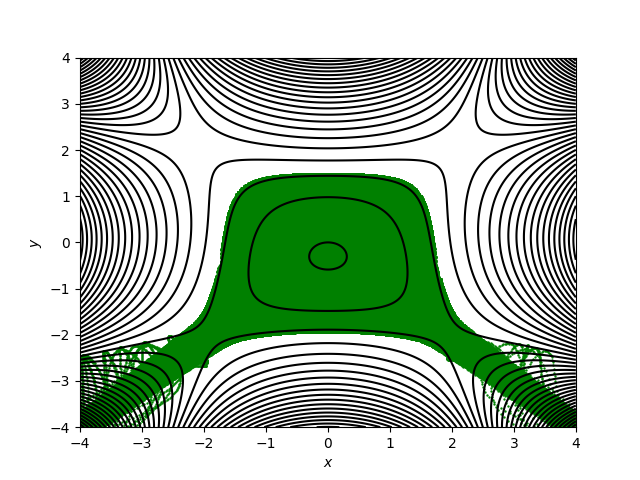}
\caption{ Upper Left Panel: Trajectories that have initial  conditions 
on the dividing surface of a periodic orbit of the family of periodic orbits of lower left hand  saddle for E=17 and contours of the potential energy.
Upper Right Panel: Trajectories that have initial  conditions 
on the dividing surface of a periodic orbit of the family of periodic orbits of the lower right  hand  saddle for E=17 and contours of the potential energy.
 Lower  Left  Panel:  Trajectories that have initial  conditions 
on the dividing surface of a periodic orbit of the family  of periodic orbits of the central minimum for E=17 and contours of the potential energy.
 Lower  Right Panel:  Trajectories that have initial  conditions 
on the dividing surface of an unstable  periodic orbit with period 10 for E=17 and contours of the potential energy.}
\label{div17}
\end{figure}

In  this subsection, we study a representative example  for the first case,  E=17. The periodic orbits of  the central minimum are stable. In  Fig. \ref{div17} we see that the trajectories that  have initial conditions on the dividing surfaces of the periodic orbits of the family of central minimum are trapped for a long time at the central region of the caldera and then they exit from the regions of the lower saddles. In addition, we observe in the same figure that  the trajectories that  have initial conditions on the dividing surfaces of the periodic orbits of families of the lower saddles are trapped  for a long time after the entrance into  the caldera and before they exit from it. 

\begin{figure}
 \centering
\includegraphics[angle=0,width=10.0cm]{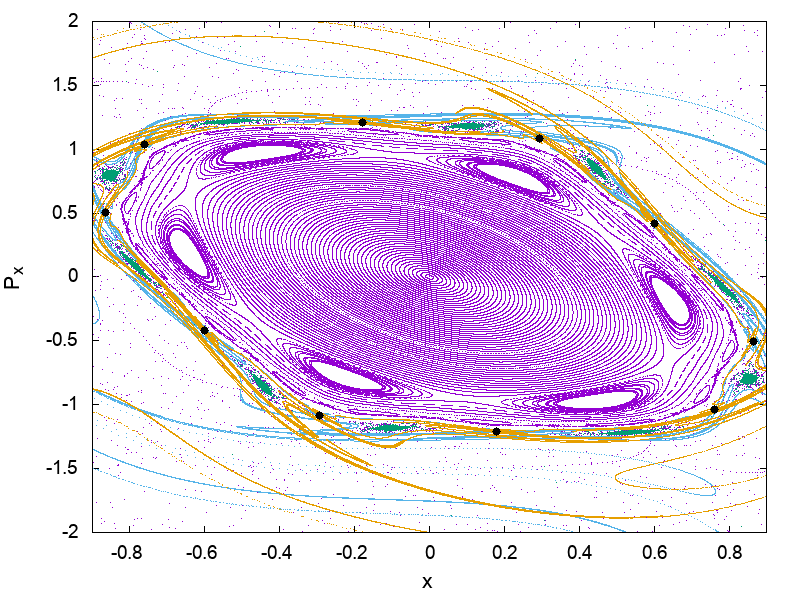}
 \includegraphics[angle=0,width=10.0cm]{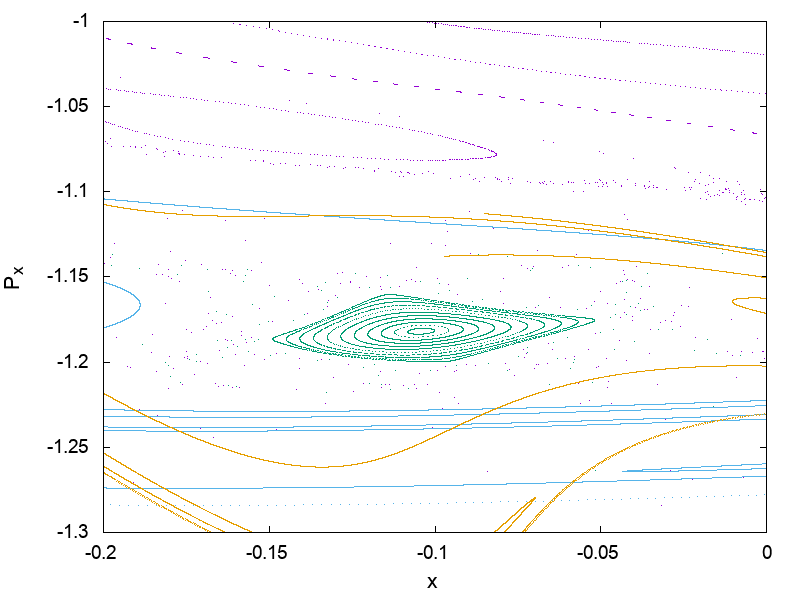}
 \caption{ Upper Panel: Poincar{\'e} surface of section $y=0$ with $p_y>0$ for E=17. We observe invariant curves around the stable periodic orbit of the family of central minimum. We also  observe the unstable (with cyan color) and stable (with orange color)  invariant manifolds of an unstable periodic orbit with period 10, that is represented by ten black points. Furthermore, we see the invariant curves (with green color) around the ten points of the accompanied stable periodic orbit with period 10.
 Lower Panel:  An enlargement of the previous figure in order to see the invariant curves (with green color)  in one of the ten stable regions that are around  the ten points of the  stable periodic orbit with period 10.}
\label{pos1y0-17}
\end{figure}

We  investigated  the origin of this trajectory  behaviour through the study of the phase space structure. We studied the phase space in the neighborhood of the stable periodic orbit of the central minimum  on the Poincar{\'e} section $y=0$ with $p_y>0$. We  see in figure \ref{pos1y0-17} invariant curves  around the  stable periodic orbit of the central minimum that is represented by one point at the center (0,0). The existence of these curves is guaranteed by the KAM (Kolmogorov-Arnold-Moser) theorem \cite{kolmo54,arn63,mos62}.  According this theorem there are trajectories that lie on 2-dimensional invariant tori (KAM invariant tori)  around the stable periodic orbits in a  Hamiltonian system with two degrees of freedom (see also \ref{sub2}). These tori are represented by invariant curves  on  the 2D Poincar{\'e} section like those that we observe around the stable periodic orbit of the central minimum in Fig. \ref{pos1y0-17}. This means that the trajectories close to the stable periodic orbit   of the central minimum  lie 
 on these tori for ever and they  are trapped in the central region of the caldera. 

 \begin{figure}
 \centering
\includegraphics[angle=0,width=8.0cm]{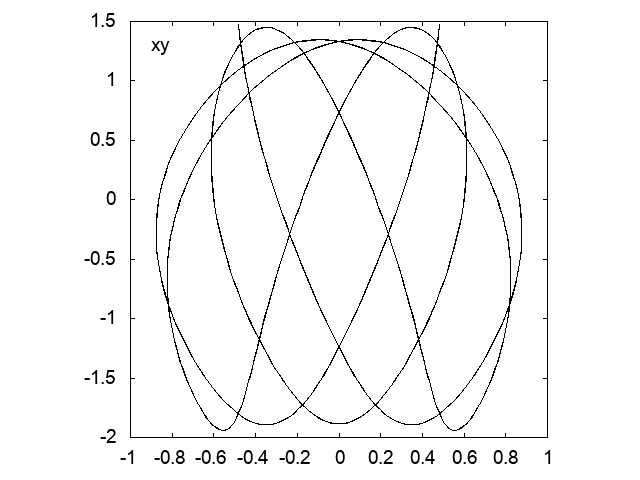} 
 \caption{An unstable periodic orbit with period 10 in the configuration space, for E=17.}
\label{per10}
\end{figure}
 
In  Fig. \ref{pos1y0-17} we see also invariant curves around a 
stable 6-periodic orbit  between the invariant curves around the central stable periodic orbit. All these trajectories  are trapped between the tori 
that are around the stable periodic orbit of the central minimum (see \ref{sub2}).  This explains that there are trajectories  around the central area of the caldera but it cannot explain the transport of  trajectories  to the regions of lower saddles.

If we observe carefully the upper panel of  Fig. \ref{pos1y0-17} we can discern ten small green regions that are outside the invariant curves .  We can see  by enlargement  one of these regions in the lower panel in the Fig. \ref{pos1y0-17} and we observe small green invariant curves. This means that  these regions are stable areas and  indicate the presence of a stable periodic orbit with period 10.  This means that we have also an accompanied unstable periodic orbit with period 10
that is represented by ten black points (upper panel of Fig. \ref{pos1y0-17}) . The morphology of this orbit in the configuration space can be seen in  Fig. \ref{per10}.

We computed the invariant manifolds of the unstable periodic orbit with period 10 on the 2D Poincar{\'e} section $y=0$ with $p_y>0$. The invariant manifolds of a periodic orbit in a Hamiltonian system system with two degrees of freedom  are 2D  geometrical objects in the energy surface   and their existence is guaranteed by the theorem of unstable and stable manifolds \cite{wig03}. These  are 1-dimensional objects on the 2D Poincar{\'e} surfaces of section. We have two kinds of invariant manifolds, the unstable manifold and the stable manifold.  All trajectories with initial conditions on these objects approach the periodic orbit (stable manifold) and  move away from the periodic orbit (unstable manifold). 

The computation of the invariant manifolds  on the 2D Poincar{\'e} section $y=0$ with $p_y>0$ has been done numerically. As we know (see appendix A) an unstable periodic orbit has  two real eigenvalues of the monodromy matrix of the Poincar{\'e}  map. The one is inside the unit circle and associated with  the stable manifold and the other is outside of the unit circle and associated with  the unstable manifold.  For the calculation of the unstable and stable manifolds we need to take many initial conditions (in this paper 20000 ) in the direction of the  unstable eigenvector (the eigenvector of the monodromy matrix of the Poincar{\'e}  map that corresponds to the eigenvalue outside the unit circle)  and stable eigenvector (the eigenvector of the monodromy matrix of the Poincar{\'e}  map that corresponds to the eigenvalue inside the unit circle)  and take their consequents and their antecedents respectively.

\begin{figure}
 \centering
\includegraphics[angle=0,width=10.0cm]{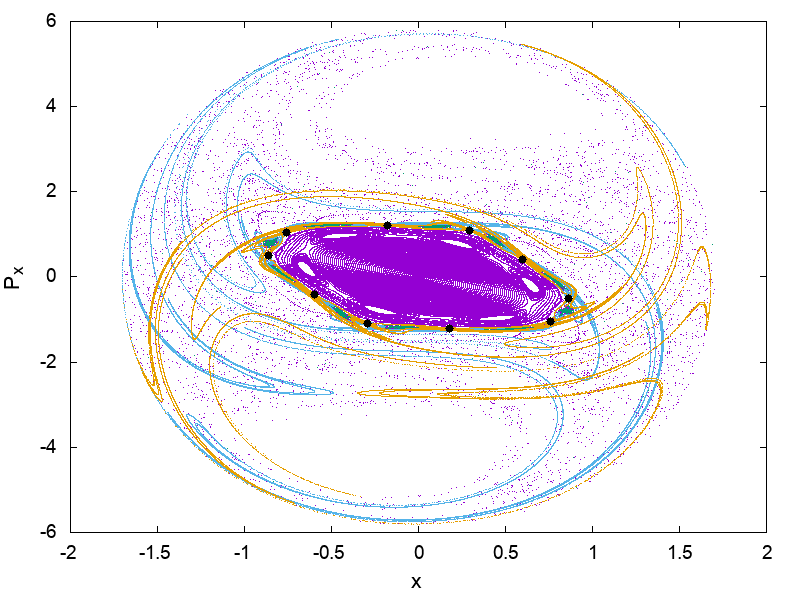} 
 \caption{ Poincar{\'e} surface of section $y=0$ with $p_y>0$ for E=17. We see the extension of the invariant manifolds
 (unstable with cyan color and stable with orange color) in the whole available space. The central region   of the figure is  depicted  by enlargement  of the upper panel  of Fig.\ref{pos1y0-17}.}
\label{pos2y0-17}
\end{figure}

The unstable and stable manifolds of the periodic orbit with period 10 are depicted by cyan and orange colors in Figs. \ref{pos1y0-17} and \ref{pos2y0-17}. We see that the invariant manifolds  start from the ten points of the periodic orbit (Fig. \ref{pos1y0-17} ) and have many oscillations, that is an indication of strong chaos. Then the manifolds extend throughout the whole available space (Fig. \ref{pos2y0-17} ). This is because of the fact that this periodic orbit is outside the   central stable region around the stable periodic orbit of the central minimum and there are not restrictions to 
the manifolds of the periodic orbit  to occupy the available space except the space that is occupied from the tori around the stable periodic orbit of the central minimum . Moreover, we can observe in Fig. \ref{pos2y0-17} that many points are trapped in the lobes of the invariant manifolds for long time. This means that there are many trajectories that are trapped by the invariant manifolds of the unstable periodic orbit with period 10.  This  mechanism explains the trapping of trajectories in the central region of the caldera and the transport of these trajectories  to large regions of the phase space.  The trajectories in the neighbourhood of an unstable periodic orbit  will  move away from the periodic orbit following  the unstable manifold. If we take initial conditions on  the dividing surface of the unstable periodic orbit with period 10, we will observe that the trajectories, after their trapping by the manifolds for long time,  exit from the regions of the lower saddles (lower right panel of Fig.\ref{div17}). From this behaviour we conclude that the unstable manifold of the unstable periodic orbit of period 10 guides the trajectories far away from the neighbourhood of the periodic orbit and to the exit from the two regions  of lower saddles  of the caldera. 

\begin{figure}
 \centering
\includegraphics[angle=0,width=10.0cm]{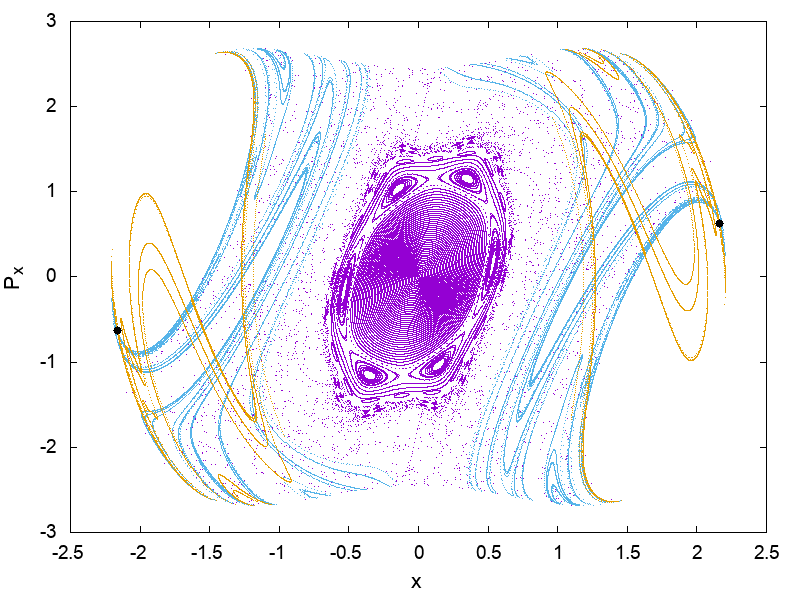} 
 \caption{ Poincar{\'e} surface of section $y=-1.8019693$ with $p_y>0$ for E=17. We see the unstable (cyan color) and stable (orange color) invariant manifolds of the periodic orbits of two lower saddles, that are represented by two black points. Furthermore, we observe invariant curves around the stable periodic orbit of the central minimum and cloudy points outside from them. }
\label{pos1ya-17}
\end{figure}

As we  described above, the invariant manifolds of the unstable periodic 
orbit with period 10 are responsible for the trapping and the transport  of trajectories in the caldera. But,  what is the role of  unstable periodic orbits of lower saddles? In order to understand this we computed the invariant manifolds of the unstable periodic orbits of the lower saddles  on the 2D Poincar{\'e} section $y=-1.8019693$ with $p_y>0$ (this is the surface of section on which we can see these periodic orbits because they do not cross the plane $y=0$). As we can see in Fig. \ref{pos1ya-17}, the invariant manifolds are depicted with cyan (for the unstable manifold) and orange (for the stable manifold). The invariant manifolds extend into the central region which is occupied from the invariant curves around the stable periodic  of the central minimum and many scattered points that are outside of them. We observe also that many points are trapped in the lobes of the invariant manifolds. This means that many trajectories are transported from the unstable  manifold of the periodic orbits of the lower saddles far away from the periodic orbits towards  the central region of the caldera. Then the trajectories are trapped by the invariant manifolds of the unstable periodic orbit with period 10  as we can see in  Fig. \ref{pos2y0-17}. The trajectories  in the neighbourhood of an unstable periodic orbit approach the periodic orbit following the stable manifold. This means that many  trajectories, that are trapped, are guided from the stable manifold to the central region of the caldera  which is the region of the unstable periodic orbit with period 10.  This  explains the fact that the trajectories that have initial conditions on the dividing surfaces of the periodic orbits of the lower saddles  enter the central region of the caldera (see  upper right and left panels of  Fig. \ref{div17}).

 \subsection{Case II}
 \label{res1b}
 
In this case the periodic orbits of the central minimum have become unstable and we have the appearance of new stable periodic orbits with period 2 after the bifurcation  of the family of the central minimum. We study a representative example  for E=23. The trajectories that have initial conditions on the dividing surfaces of the periodic orbits of the family of the central minimum, its bifurcation  and families of lower saddles have similar behaviour like this that we observed in the case I (see Fig. \ref{div17}). 

\begin{figure}
 \centering
\includegraphics[angle=0,width=10.0cm]{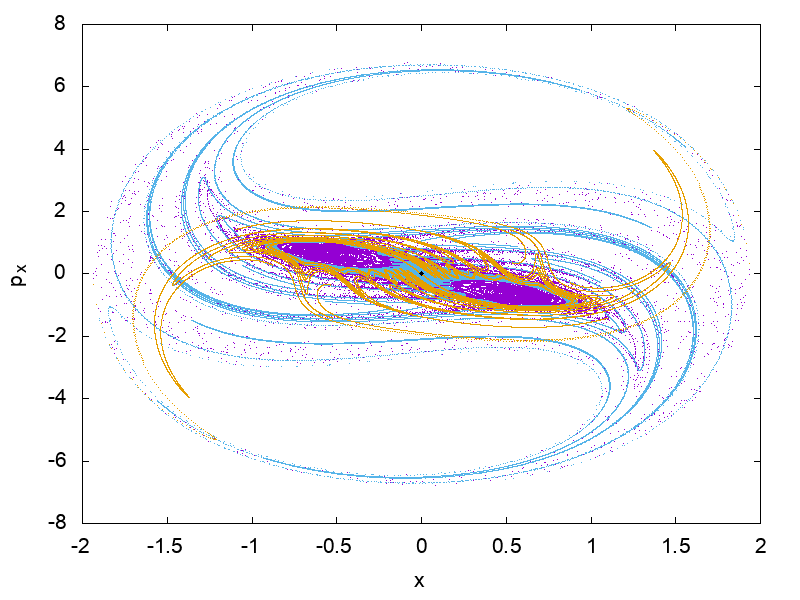}
 \includegraphics[angle=0,width=10.0cm]{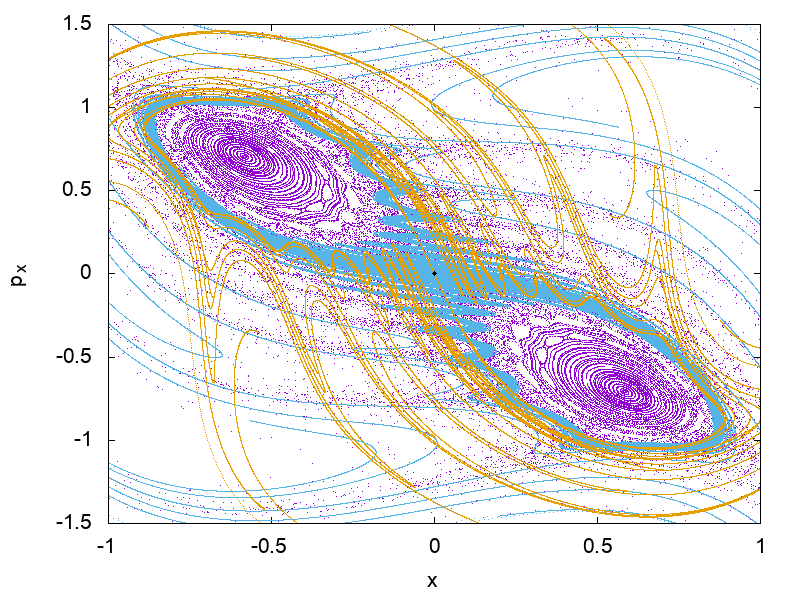}
 \caption{ Upper Panel: Poincar{\'e} surface of section $y=0$ with $p_y>0$ for E=23. We observe invariant curves around the two points of the stable periodic orbit with period 2 of the bifurcation of the central minimum. We also  observe the unstable (with cyan color) and stable (with orange color)  invariant manifolds of the unstable periodic orbit of the family of the central minimum (the black point  at the center). 
 Lower Panel:  An enlargement of the previous figure in order to see the central region  around the periodic orbit of the central minimum and the invariant curves around the stable periodic orbit with period 2 of the bifurcation of the  family of the central minimum }
\label{pos1y0-23}
\end{figure}

In order to study the origin of this trajectory  behaviour we studied the phase space structure for E=23. In Fig. \ref{pos1y0-23}, we see that there  are no invariant curves, on the Poincar{\'e} section $y=0$ with $p_y>0$, around the periodic orbit of the family of the central minimum, but scattered collection of points. Moreover, we observe invariant curves around a stable periodic orbit with period 2.  This periodic orbit  belongs to the bifurcation of the central family of the central minimum (lower panel of  Fig.\ref{pos1y0-23}). We have many trajectories that lie on the tori in the phase space (invariant curves  on the Poincar{\'e} section) and they are trapped forever in the central region around the stable periodic orbit with period 2. Furthermore, we observe many points on the Poincar{\'e} section (Fig. \ref{pos1y0-23}) inside the invariant manifolds of the unstable periodic orbit of the central minimum. This has as a consequence that many trajectories  are trapped by the invariant manifolds of the periodic orbit of the central minimum and they are transported from the unstable invariant manifold far away from the periodic orbit and finally  to  the  exit from the caldera. This is the same  mechanism as described in case I with the only difference that the unstable periodic orbit belongs to the family of the central minimum and it does not belong to a family of periodic orbits with period 10, as in  case I. 

   \begin{figure}
 \centering
\includegraphics[angle=0,width=10.0cm]{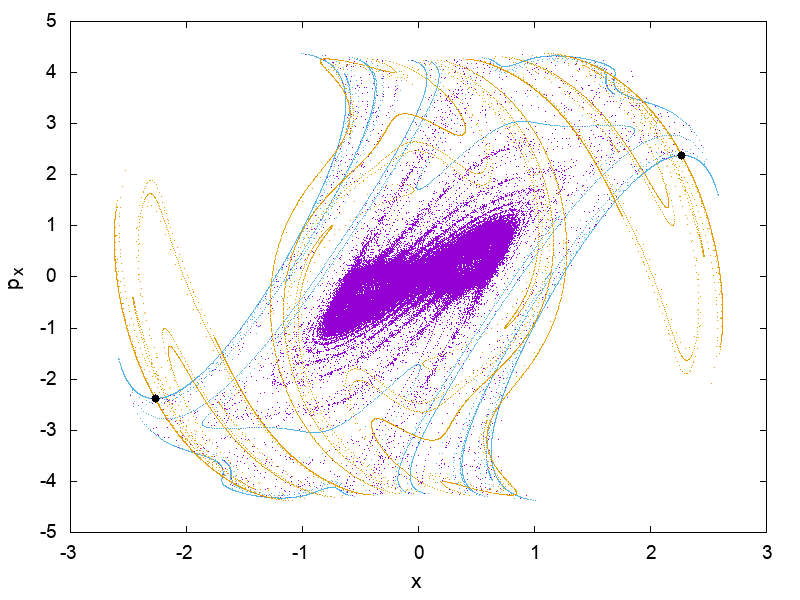} 
\includegraphics[angle=0,width=10.0cm]{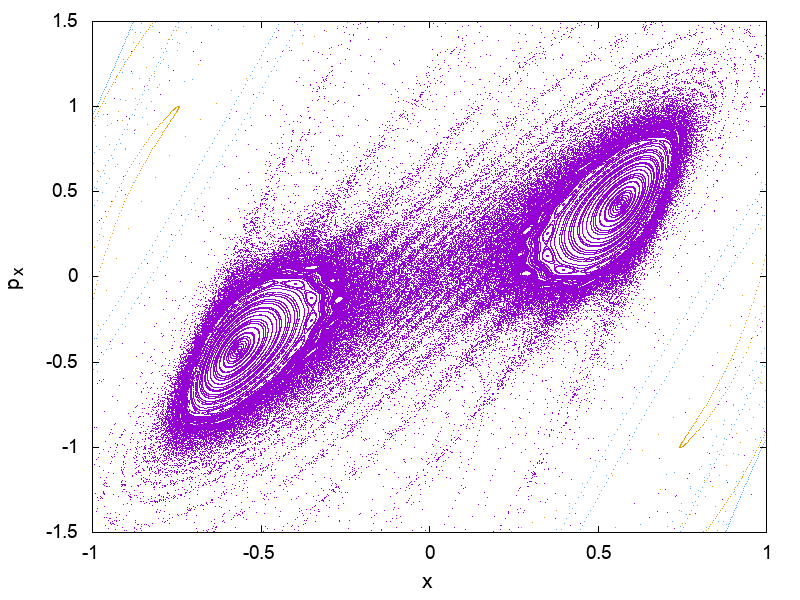} 
 \caption{Upper panel:  Poincar{\'e} surface of section $y=-1.8019693$ with $p_y>0$ for E=23. We see the unstable (cyan color) and stable (orange color) invariant manifolds of the periodic orbits of the two lower saddles, that are represented by two black points. Furthermore, we observe invariant curves around the stable periodic orbit with period 2 of the bifurcation of the  family of the central minimum and scattered collection of points outside from them. 
 Lower panel: An enlargement of the previous figure in order to see the central region  around the periodic orbit of the central minimum and the  invariant curves around the stable periodic orbit with period 2 of the bifurcation of the  family of the central minimum.}
\label{pos1ya-23}
\end{figure}      

We also study the phase space structure close to periodic orbits of the lower saddles.  In  Fig. \ref{pos1ya-23}, we observe  similar behaviour close to the neighbourhood of the periodic orbits of the lower saddles, on the Poincar{\'e}  section $y=-1.8019693$ with $p_y>0$,  with the behaviour that we encountered  close to the neighbourhood of the periodic orbits of lower saddles in  case I. The only difference  is the central region. We have no invariant curves around the central periodic orbit of the central minimum as in fig. \ref{pos1ya-17}, but we have invariant curves around the periodic orbit with period 2 of the bifurcation of the family of the central minimum (lower panel of the Fig. \ref{pos1ya-23}).
We noticed the fact that the phase space at the neighbourhood of the periodic orbits of the lower saddles has not changed  because these families have no  bifurcations that would change the structure of  the phase space in the neighbourhood of their periodic orbits. The change of the phase space happens at the central region which is  in the neighbourhood of the periodic orbits of the central minimum. This means that the mechanism of the entrance of the trajectories into the caldera does not change.  It is the same as the corresponding mechanism in  case I. The trajectories are trapped by the invariant manifolds  of the periodic orbits of the lower saddles and they are guided by the unstable invariant manifolds close to the central region. Then they are trapped  by the invariant manifolds  of the periodic orbits  of the family of the central minimum (Fig. \ref{pos1y0-23}) and they are transported  from the stable manifolds to the center of the caldera.

\subsection{Case III}
\label{res1c}

\begin{figure}
 \centering
\includegraphics[angle=0,width=5.0cm]{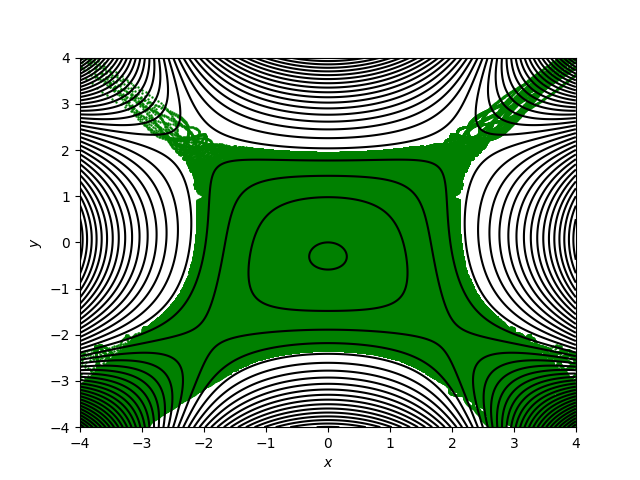}
 \includegraphics[angle=0,width=5.0cm]{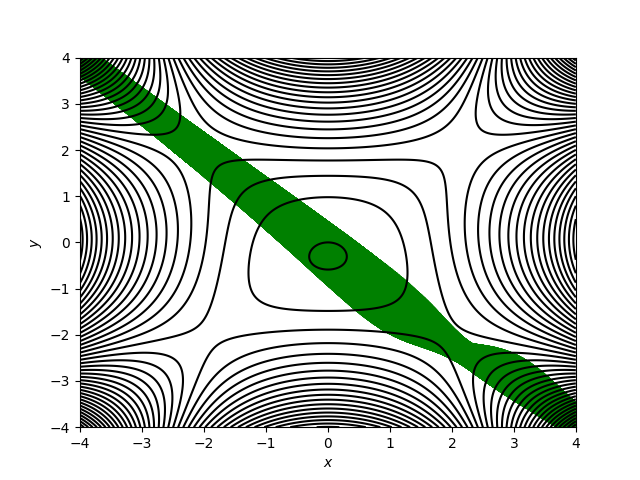}
 \includegraphics[angle=0,width=5.0cm]{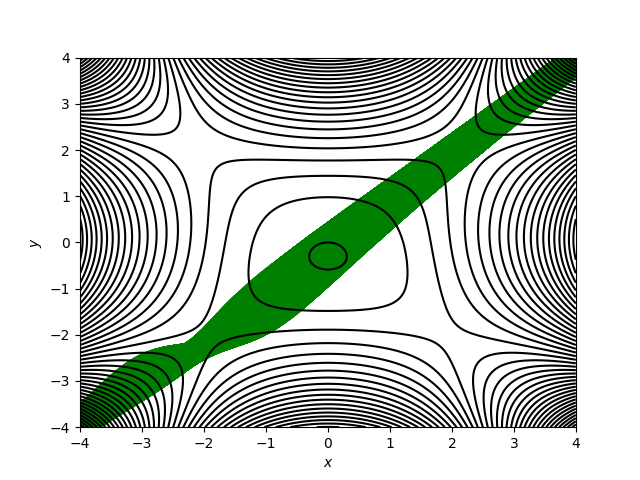}\\
\caption{  Left Panel:  Trajectories that have initial  conditions 
on the dividing surface of a periodic orbit of the family  of periodic orbits of the central minimum for E=29 and contours of the potential energy.
Central  Panel: Trajectories that have initial  conditions 
on the dividing surface of a periodic orbit of the family of periodic orbits of the upper left hand  saddle for E=29 and contours of the potential energy.
Right Panel: Trajectories that have initial  conditions 
on the dividing surface of a periodic orbit of the family of periodic orbits of the upper right  hand  saddle for E=29 and contours of the potential energy.}
\label{div29}
\end{figure}

In  case III we have a similar situation as in  case II with the difference being that we have the appearance of the families of periodic orbits of the upper saddles. In this subsection we study a representative example for E=29. The trajectories that have initial conditions on the dividing surfaces  of the periodic orbits of the central minimum are trapped for a long time at the central region of the caldera and then they exit from the four regions of the saddles (Fig. \ref{div29}). The trajectories that have initial conditions on the dividing surfaces of the periodic orbits of the families of the upper saddles  go straight across  the caldera and exit via the  opposite lower saddle (Fig. \ref{div29}). This phenomenon is known as "dynamical matching"  \cite{col14}.

\begin{figure}
 \centering
\includegraphics[angle=0,width=10.0cm]{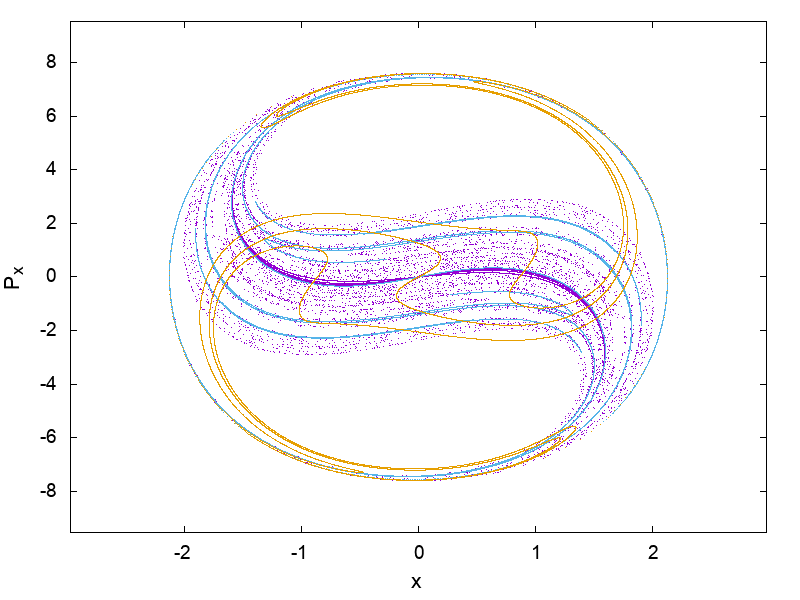} 
 \caption{ Poincar{\'e} surface of section $y=0$ with $p_y>0$ for E=29. We see the invariant manifolds (unstable with cyan color and stable with orange color) of the unstable periodic orbit of the family of the central minimum.}
\label{pos1y0-29}
\end{figure}

The mechanism of the trapping of  trajectories  and exit  from the caldera is the same as the corresponding mechanism  in  case II. As we can see in  Fig. \ref{pos1y0-29} the points of the trajectories are trapped inside the invariant manifolds of the periodic orbit of the family of the central minimum. This means that, as in  case II, the invariant manifolds of the periodic orbits of the central minimum are responsible for the trapping and the transport of the trajectories to the  exit from the caldera (through the regions of the four saddles - see Fig. \ref{div29}). 

The mechanism of the entrance of the trajectories from the lower saddles 
to the central region of the caldera is the same as described in the previous subsections (cases I and II). This is because there is no change of the phase space structure at the neighbourhood of the periodic orbits of the families of the lower saddles.  The trajectories  that  begin from the dividing surfaces of the periodic orbits of the families of the lower saddles have the same behaviour as that shown in the left panel of the Fig. \ref{div29}. The only difference is that we have fewer trajectories  than those of the left panel of the  Fig. \ref{div29} to exit through the regions of the upper saddles. The reason is that the trajectories are trapped two times, one time by the invariant manifolds of the periodic orbits of the lower saddles and one time by the invariant manifolds of the periodic orbit of the family of the central minimum, for our time scale.

 \begin{figure}
 \centering
\includegraphics[angle=0,width=10.0cm]{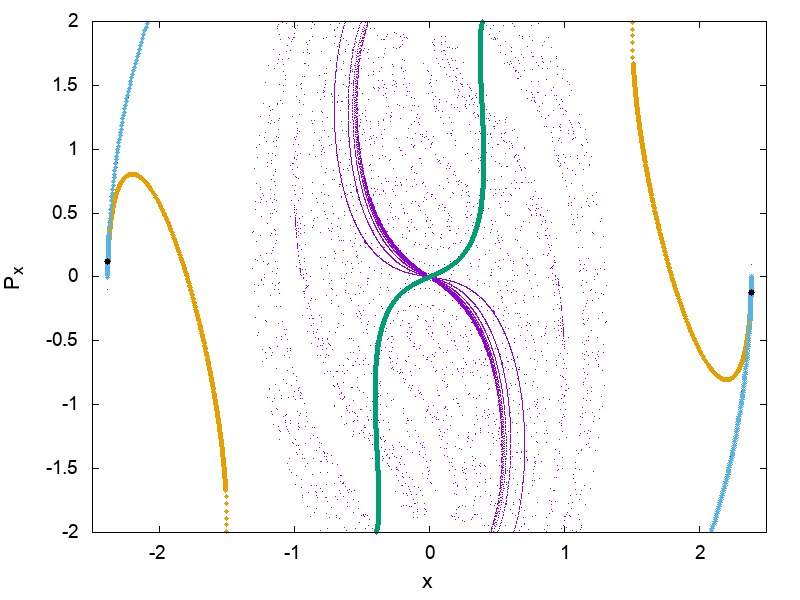} 
 \caption{ Poincar{\'e} surface of section $y=1.884090$ with $p_y>0$ for E=29. We see the unstable (cyan color) and stable (orange color) invariant manifolds of the periodic orbits of the two upper saddles, that are represented by two black points. Furthermore, we observe the invariant manifolds (unstable with violet color and stable with green color) of the periodic orbit   of the family of the central minimum.}
\label{pos1ya-29}
\end{figure}

Next we study the phase space structure close to the periodic orbits 
of the families of the upper saddles using the Poincar{\'e} section $y=1.884090$ with $p_y>0$ (this is the section for which we can see these periodic orbits because they do not cross the plane $y=0$).  We can observe in Fig. \ref{pos1ya-29} that the  central region has  invariant manifolds of the periodic orbit of the family of the central minimum and scattering points around  it. But the invariant manifolds of the periodic orbits of the families of the upper saddles are far away from the central region. The stable manifolds (with orange color in  Fig. \ref{pos1ya-29})  bring  trajectories at the neighbourhood of the periodic orbits that were far away from the central region and the unstable manifolds guide 
(with cyan color in  Fig. \ref{pos1ya-29}) the points to  infinity and far away from the central region. This means that we have no interaction of the invariant manifolds with the central region of the caldera and this explains the dynamical matching that we observe in the configuration space. Also the non-existence of the interaction of the invariant manifolds with the central region of the caldera explains the reason that we have fewer trajectories to leave from the upper saddles in the left panel of the Fig. \ref{div29}. This happens because the trajectories  that are trapped by the invariant manifolds of the unstable periodic orbit of the family of the central minimum can be trapped inside the invariant manifolds of the periodic orbits of the families of the lower saddles, as we can see in  Figs. \ref{pos1ya-17} and \ref{pos1ya-23} for the cases I and II (this mechanism is the same for the case III). After this the trajectories are guided
 from the stable manifolds to the neighbourhood of the periodic orbits of the lower saddles. Moreover, the invariant manifolds of the periodic orbits of the upper saddles do not interact with the manifolds of the periodic orbits of the central region that are responsible for the transport of the trajectories to the region of the upper saddles.

\subsection{Case IV}
\label{res1d}

In  case IV, we investigated a representative example for E=30. 
In this case the periodic orbits of the family of the central minimum have become stable again, after the second 
bifurcation of the family of the central minimum. We have also  a new family of unstable periodic orbits with period 2 that belongs to the bifurcation  of the family of the central minimum.  The trajectory behaviour of the orbits that have initial conditions on the dividing surfaces of the periodic orbits of the family of the central 
minimum and its bifurcation is the same as the trajectory  behaviour that is represented in the left panel of the Fig. \ref{div29} in  case III. The behaviour of trajectories that begin from the dividing surfaces of the periodic orbits of the families of the saddles is similar to that of the corresponding trajectories  in  case III. 
 \begin{figure}
 \centering
 \includegraphics[angle=0,width=10.0cm]{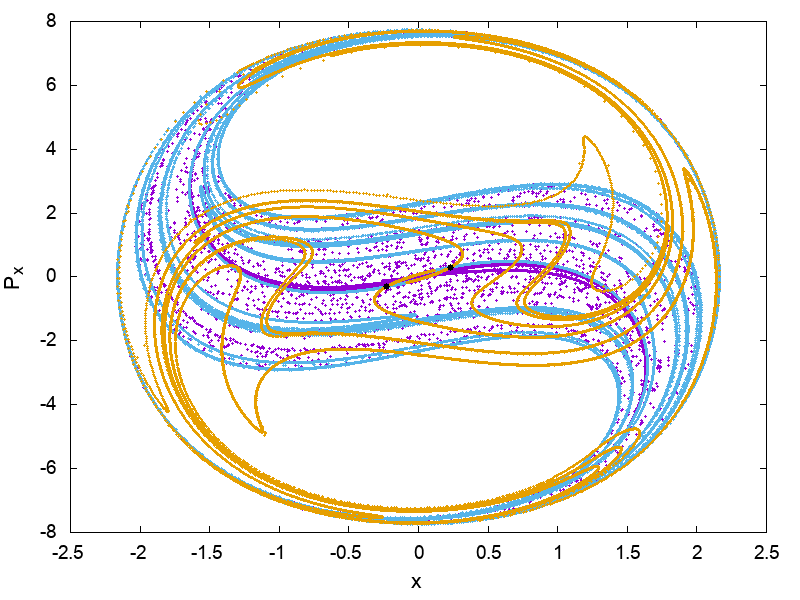}		 
\includegraphics[angle=0,width=10.0cm]{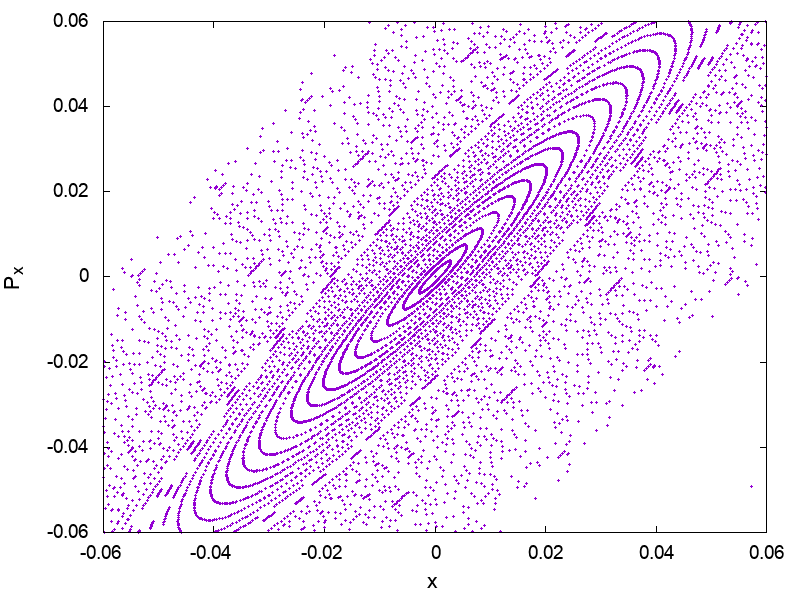} 
 \caption{Upper panel:  Poincar{\'e} surface of section $y=0$ with $p_y>0$ for E=30. We see the invariant manifolds (unstable with cyan color and stable with orange color) of the unstable periodic orbit  with period 2 (that is represented by two black points) of the bifurcation of the family of the central minimum.
Lower panel:  An enlargement of the figure of the upper panel in order to see the invariant curves around the periodic orbit of the central minimum.}
\label{pos1y0-30}
\end{figure}

We  observe that the phase space structure close to the periodic orbit
of the central minimum, on the Poincar{\'e} section $y=0$ with $p_y>0$ 
( Fig. \ref{pos1y0-30}), is similar to the phase space structure close to the periodic orbit of the same family in  case I. We observe invariant curves around the stable periodic orbit of the central minimum and  many trajectories that are trapped  by the invariant manifolds of the unstable periodic orbit with period 2 of the bifurcation of the family of the central minimum. The only difference is that  we have now an unstable periodic orbit with period 2, while in case I we have an unstable periodic orbit with period 10.  This means  that  the mechanism of trapping and guiding of the trajectories  is the same as case I. The invariant manifolds of the unstable periodic orbit with period 2 are responsible for the trapping of trajectories in the central region of the caldera and their transport to the exit  from the caldera.

The structure of the phase space in the neighbourhood of the periodic orbits of the families of the  lower  saddles has not changed. This is  expected because we have no  further bifurcation or other dynamical phenomenon for these families. This has as a result that the mechanism for the entrance into the caldera  from the region of the lower saddles will be the same as for  the previous cases I,II and III. This means that the  trajectories  are trapped by the invariant manifolds of the unstable periodic orbits of the lower saddles and they follow the unstable manifold of these periodic orbits close to the central region of the caldera. Then the trajectories are trapped again by the invariant manifolds of the  unstable periodic orbits with period 2 of the bifurcation of the family of the central minimum and they follow the stable manifolds of these unstable periodic orbits to the central region of the caldera.

Similar to the case of the lower saddles, the phase space in the neighbourhood of the unstable periodic orbits of the upper saddles does not change. This means that the invariant manifolds do not interact with the central region of the caldera as in  case III. This results in the `dynamical matching' (see  \ref{res1c})  of the trajectories that begin on the dividing surfaces  of the periodic orbits of the families of upper saddles.

\newpage

\section{Summary and conclusions}
\label{sec:summary}
We studied the phase space structure and transport of the 2D caldera model. We encountered four different cases of the phase space  structure. The phase space structure is determined by the transition of the stability of the family of periodic orbits of the central minimum from stability to instability and vice versa. Moreover, it is determined by the appearance of the new families of periodic orbits, that are bifurcations of the family of the central minimum, and the value of the energy. These cases are described in detail in \ref{res}. 

In this paper, we investigated the origin  of the trajectory behaviour of the Caldera model and we found three mechanisms for this. These  mechanisms are based on certain geometrical objects of the phase space, the invariant manifolds of the unstable periodic orbits and the KAM invariant tori. We saw that the geometry of the phase space give us important information for the trajectory behaviour that cannot be obtained from the study of the configuration space and the use of statistical methods.  The  mechanisms for the trajectory behaviour are:
\begin{enumerate} 
\item {\bf The trapping of  trajectories and the entrance into the Caldera}: 
The trajectories that have initial conditions  on the dividing surfaces of the unstable  periodic orbits of the low energy saddles are trapped inside the invariant manifolds  of the unstable periodic orbits and they follow the  unstable manifold (far away from the periodic orbits of the lower saddles), until they are trapped 
from the invariant manifolds of the unstable periodic orbits of the central region. The periodic orbits of the central region are, the unstable periodic orbits with period 10 that are outside  the stable region of the  stable  periodic orbits of the family of the central minimum (case I),  the unstable periodic orbits of the family of the central minimum (cases II and III) and  the unstable periodic orbits with period 2 of a period-doubling bifurcation of the family of the central minimum (case IV). Then the trajectories follow the stable manifolds of these unstable periodic orbits towards to the central region of the caldera. 

\item {\bf The trapping of trajectories  and the exit from the Caldera}: 
The trajectories that have initial conditions  at the central region of the potential are divided into two categories.
In the first category, the trajectories are trapped for ever because they lie on or  are inside the KAM invariant tori surround  the stable periodic orbits that exist at the central region of the potential. In the second category, the trajectories are trapped inside the lobes of the invariant manifolds  of the unstable periodic orbits of the  central region (these periodic orbits are described in the first mechanism). The trajectories are transported from the unstable manifolds of these periodic orbits, via the regions of the saddles, to the exit from the caldera. 

There is also a secondary mechanism that supports the basic mechanism of the exit from the caldera. 
 In this secondary mechanism, many trajectories that are guided by the unstable manifolds of the periodic orbits of the central region are trapped inside the invariant manifolds of the unstable periodic orbits of the lower energy saddles. Afterwards, the trajectories  follow the stable manifolds of the periodic orbits of the lower energy saddles, via the regions of the lower saddles,  to the exit from the caldera. This secondary mechanism does not exist for the high energy saddles because the invariant manifolds of the unstable periodic orbits of the high energy saddles are not connected with the central region of the potential. This means that this secondary mechanism supports  the exit of the trajectories  from the caldera via the regions  of the lower energy saddles. This is the reason why the majority of the trajectories exit from the regions of the lower energy saddles.

\item {\bf Dynamical matching}: The trajectories  that have initial conditions on the dividing surfaces of the periodic orbits of the high energy saddles go straight across the caldera and exit via the opposite lower saddle (dynamical matching \cite{col14}). This behaviour is due to the fact that the invariant manifolds of the unstable periodic orbits of the high energy saddles do not interact  with the central region of the caldera. The trajectories  follow the invariant manifolds far away from the central region of the caldera, without any interaction with it.

Although the phase space structures described in this paper were derived for a 2D potential, the experimental observations for the reactions cited in the introduction suggest that broadly similar objects may exist in the high dimensional phase spaces representing the reactions of polyatomic molecules occurring on caldera-type potentials.
\end{enumerate}

\nonumsection{Acknowledgments} We acknowledge the support of EPSRC Grant No.~EP/P021123/1 and ONR Grant No.~N00014-01-1-0769.



\appendix{Linear stability and bifurcations of periodic orbits}
\label{appen}
We  present the method of the computation of the linear  stability of  periodic orbits. The 2D Poincar{\'e} map  of a Hamiltonian system of two degrees of freedom (like what is considered in this paper) is given by  equations of the form :

\begin{eqnarray}
x_1=f_1(x_0,p_{x_0})
\nonumber\\
p_{x_1}=f_2(x_0,p_{x_0})
\nonumber\\
\vdots
\nonumber\\
x_n=f_1^{n}(x_0,p_{x_0})
\nonumber\\
p_{x_n}=f_2^{n}(x_0,p_{x_0})
\nonumber
\end{eqnarray}

For a periodic orbit with period n we have  $x_n=x_0$ and $p_{x_n}=p_{x_0}$. The functions  $f_1$ and $f_2$ are obtained by numerical integration of an trajectory with initial conditions   on the  Poincar{\'e} section until its first return  to  the  Poincar{\'e} section. The functions  $f_1^{n}$ and $f_2^{n}$ are obtained by numerical integration of an trajectory with initial conditions   on the  Poincar{\'e} section until  its  nth return  to  the  Poincar{\'e} section. 

We have two types of periodic orbits, according to their linear stability: stable and unstable periodic orbits. The numerical calculation of the linear  stability of periodic orbits in 2D autonomous Hamiltonian systems is based on the method of H{\'e}non (see e.g.  \cite{conto02}). We first consider small deviations from the initial conditions of the periodic orbit on the  2D Poincar{\'e} surface of section and then  integrate the trajectory again to the next  intersection. The relation of the final deviations of this neighbouring trajectory (after the first return to the  Poincar{\'e} surface of section)  from the periodic orbit, with the initially introduced deviations can be written in vector form as $\xi=M\xi_0$, where $\xi$ is the final deviation (after the first return to the  Poincar{\'e} surface of section) and $\xi_0$  is the initial deviation, M is a 2$\times$2 matrix, called the monodromy  matrix. 

\begin{equation}
M=
 \begin{bmatrix}
a & b\\
c & d
\end{bmatrix}
  \end{equation}

This matrix satisfies the symplectic identity\cite{arn78}:  $M^{T}JM=J$ where $M^{T}$ denotes the transpose of $M$  and $J$ is the fixed  nonsingular, skew-symmetric 2$\times$2  matrix:

\begin{equation}
J=
 \begin{bmatrix}
0 & 1\\
-1 & 0
\end{bmatrix}
  \end{equation}
 $J$ satisfies the relation  $J^{-1}=J^{T}=-J$, where $J^{T}$ and $J^{-1}$ are the transpose and  the inverse of matrix $J$, respectively. The  symplectic identity in our case (in which we have a symplectic matrix  2$\times$2) is equivalent with the condition $Det M =1$ (where $Det$ denotes the determinant of a matrix),  $M^{T}JM=J \Leftrightarrow DetM=1$ (this is not true for the general case of a symplectic matrix $M$, that is a 2$n\times$2$n$ matrix with $n\in \mathbb{N}_{\ne 0} $, for $n>1$). The elements of the monodromy matrix $M$ are:   

\begin{eqnarray}
a=\frac{\partial f_1}{\partial x}(x_0,p_{x_0})
\nonumber\\
b=\frac {\partial f_1}{\partial p_x}(x_0,p_{x_0})
\nonumber\\
c=\frac{\partial f_2}{\partial x}(x_0,p_{x_0})
\nonumber\\
d=\frac {\partial f_2}{\partial p_x}(x_0,p_{x_0})
\end{eqnarray}

It can  be shown   that the characteristic equation  of $M$ is  $\lambda^2-(a+d)\lambda+1=0$ and the eigenvalues $\lambda_1,\lambda_2$ are the roots of this  equation. We define the quantity 
$\alpha=|a+d|/2$  to be the H{\'e}non stability parameter (see e.q. \cite{conto02}).  We have two kinds of periodic orbits:
\begin{enumerate}
\item {\bf Stable periodic orbits}: In this case  $\alpha<1$ and the eigenvalues of the monodromy matrix of a 2D Poincar{\'e} map are complex conjugate  and they are on the unit circle.
\item {\bf Unstable periodic orbits}: In this case $\alpha>1$ and the eigenvalues  of the monodromy matrix of a 2D Poincar{\'e} map are real and outside  of the unit circle.
\end{enumerate}

\begin{figure}
 \centering
\includegraphics[angle=0,width=12.0cm]{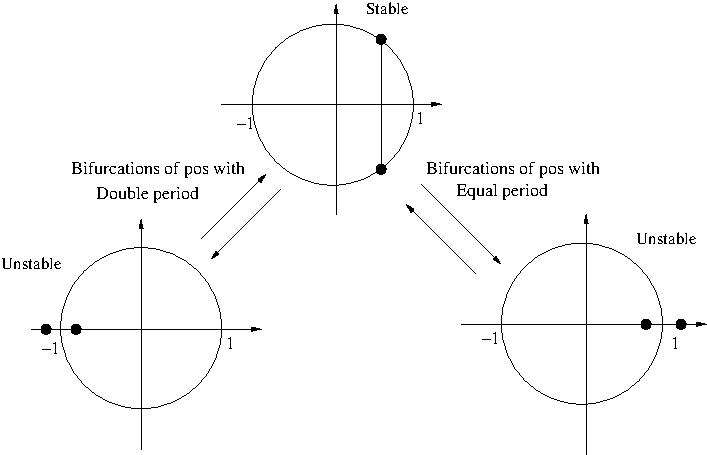}
\caption{ The Eigenvalues move on the unit circle (stable periodic orbit)   and collide at the points $\lambda=1$ and $\lambda=-1$, then the eigenvalues move on the real axis (unstable periodic orbit)
 and vice versa. The result of the  collision of the eigenvalues at the points  $\lambda=1$ and $\lambda=-1$ is that new families of periodic orbits (pos) with equal period and with double period appear in the system
 respectively.}
\label{unit}
\end{figure}

\begin{figure}
 \centering
\includegraphics[angle=0,width=6.0cm]{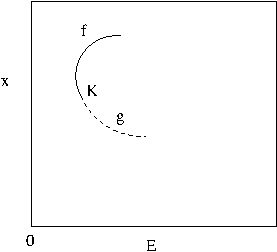}
\caption{ Characteristic near a saddle-node bifurcation at K. The solid and dashed lines depict the stable 
and unstable parts of the families of periodic orbits.}
\label{s-d}
\end{figure}

\begin{figure}
 \centering
\includegraphics[angle=0,width=5.0cm]{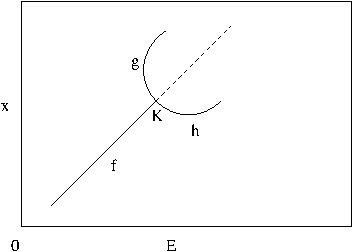}
 \includegraphics[angle=0,width=4.8cm]{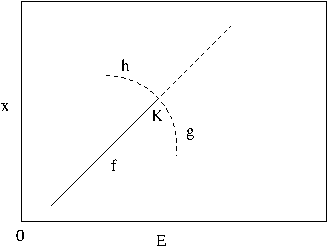}
 \caption{ Characteristic  near a pitchfork bifurcation at K. The solid and dashed lines depict the stable 
and unstable parts of the families of periodic orbits. Left panel: Supercritical pitchfork bifurcation. 
Right panel: Subcritical pitchfork bifurcation.}
\label{pitchfork}
\end{figure}

\begin{figure}
 \centering
\includegraphics[angle=0,width=5.0cm]{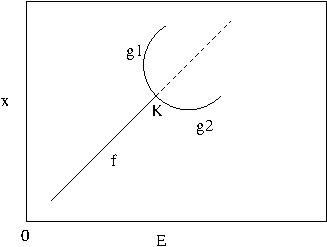}
 \includegraphics[angle=0,width=4.9cm]{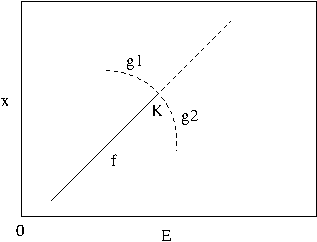}
 \caption{Characteristic  near a period-doubling bifurcation at K. The solid and dashed lines depict 
 the stable and unstable parts of the families of periodic orbits. Left panel: Supercritical period-doubling
  bifurcation. Right panel: Subcritical  period-doubling  bifurcation.}
\label{period-doubling}
\end{figure}

A family of periodic orbits  can be stable or unstable  if its periodic orbits are stable or unstable, respectively. The eigenvalues of the periodic orbits move on the unit circle (for stable families - Fig. \ref{unit}) or on the real axis (for unstable families - Fig. \ref{unit}) as a parameter of the system varies
 (in our case the parameter is the value of the Energy). If the eigenvalues meet each other at the points   $\lambda= 1$ and $\lambda= -1$ we have a collision. After this collision, the eigenvalues either move on the real axis (if they were on the unit circle before collision) or on the unit circle (if they were on the real axis before collision). This means that we have a transition of this family from stability to instability or vice versa. If the collision is at the point 
 $\lambda=1$ we have  the appearance of one or two new families of periodic orbits (pos) with equal period.  If the collision is at the point $\lambda=-1$ we have  the appearance of  two new families of periodic orbits (pos) with twice the period. These families are referred to as  bifurcations of the mother family (the initial family). There are three types of  bifurcations of equal period and  doubled period. We will describe all these types of bifurcations using the characteristic. The characteristic is a  curve that gives the initial conditions  (x or $p_x$ for our case)  of the periodic orbits of a family, on the Poincar{\'e} surface of section, as a function of a parameter of the system (Energy E in our case). 
The three types of  bifurcations are:
 
 \begin{enumerate} 
 \item  {\bf Saddle-node Bifurcation} : In this case, the characteristic  has a minimum or maximum ( for example the  minimum at K in  Fig. \ref{s-d}). The one branch of the characteristic represents  a stable family  (for example the family f  in Fig. \ref{s-d}) and  the other  represents an unstable family  (for example the family g  in Fig. \ref{s-d}). 
 
\item  {\bf Pitchfork  Bifurcation (supercritical or subcritical)} : In this case, we have two new symmetric families of equal period (g and h in  Fig. \ref{pitchfork}), that are born at the bifurcation point 
(in our example point K - see Fig. \ref{pitchfork}) and they  exist  on the side of the unstable part of the  initial family (family f in  Fig. \ref{pitchfork}) in the supercritical case ( see left panel of Fig. \ref{pitchfork}) or on the side of the  stable part of the  initial family  (family f in  Fig. \ref{pitchfork}) in the subcritical case  (see right panel of Fig. \ref{pitchfork}).  

\item {\bf Period-doubling Bifurcation (supercritical or subcritical)} : In this case, we have two  symmetric arc of a new family of double  period (g1 and g2 in  Fig. \ref{period-doubling}), that are generated at the bifurcation point (in our example point K - see Fig. \ref{period-doubling}) and they  exist  on the side of the unstable part of the  initial family (family f in  Fig. \ref{period-doubling}) in the supercritical case ( see left panel of Fig. \ref{period-doubling}) or on the side of the  stable part of the  initial family  (family f in  Fig. \ref{period-doubling}) in the subcritical case  (see right panel of Fig. \ref{period-doubling}).  
\end{enumerate}
In rare cases we have bifurcations of  families of equal or doubled period, that are similar to the Pitchfork  Bifurcation or  Period-doubling Bifurcation \cite{conto02}. The only difference is that the new families are asymmetric. This means that the two characteristics of the new families of equal period   are not symmetric to each other or the characteristic of the new family of doubled period has not two symmetric arc. These bifurcations are referred to as  asymmetric bifurcations of equal period (supercritical or subcritical) and period-doubling  asymmetric bifurcations (supercritical or subcritical).

\clearpage

\appendix{Dividing Surfaces}
\label{divsur}
In this paper we use the notion of the dividing surface (DS). A dividing surface is a surface of one less dimension than the energy surface.  A dividing surfaces is a 2D surface (with fixed total Energy) 
where the periodic orbit forms the 1D boundary of the dividing surface.  We define the dividing surface in order to divide a  region of interest, according the topography of the potential. In this  paper, the dividing surfaces are used to monitor the exit and the entrance of the trajectories to the regions of the periodic orbits of the families of four  saddles, that are connected with the entrance  into and exit from the caldera. Moreover we use the dividing surface in order to monitor the entrance into and exit from the region of periodic orbits of the families of the central minimum  and its bifurcations and  also  to the regions of periodic orbits with  period 10 that are in the central region of the potential. The main reason we choose these dividing surfaces is that we want to study, firstly the behaviour of the trajectories when entering and exiting the caldera,  and secondly the behaviour of the trajectories when entering to and leaving from the central area of the caldera. This will help us to understand and categorize the different possible paths followed by the trajectories from the central area to the  entrance and exit areas and vice versa.

The algorithm for the construction of a dividing surface (for fixed total energy E)  is given  in \cite{car18}. This  provides a discrete version of a 2D surface having the no-recrossing property and minimal flux \cite{Mau16}:
\begin{enumerate} 
\item Locate a periodic orbit.
\item Project the periodic orbit into configuration space.
\item Choose points on that curve $(x_i,y_i)$ for $i=1,...N$ where  $N$ is the desired number of points. 
Points are spaced uniformly according to distance along the periodic orbit. 
\item For each point  $(x_i,y_i)$ determine $p_{x max,i}$ by solving:
\begin{eqnarray}
\label{eq3a}
 H(x_i, y_i, p_x, 0)=\frac {p_x^2}{2m}+ V(x_i, y_i)
\end{eqnarray}
for $p_x$. Note that solution of this equation requires $E-V(x_i,y_i) \geq 0$ and there will be two solutions, $\pm p_{x max,i}$. 
\item For each point $(x_i,y_i)$ choose points for $j=1,...,K$ with $p_{x_1}=0$ and  $p_{x_K}=p_{x max,i}$  and solve the equation  $H(x_i, y_i, p_x, p_y)=E$ to obtain $p_y$.

\end{enumerate}

The time of integration of the trajectories, on dividing surfaces, was 5 time units that is  enough to see all the different kinds of trajectory behaviour \cite{col14}.


\end{document}